\documentclass[10pt,conference]{IEEEtran}
\usepackage{cite}
\usepackage{amsmath,amssymb,amsfonts}
\usepackage{algorithmic}
\usepackage{graphicx}
\usepackage{textcomp}
\usepackage{xcolor}
\usepackage[hyphens]{url}
\usepackage{fancyhdr}
\usepackage{hyperref}

\begin{document}

\title{Scalable Machine Learning Training Infrastructure for Online Ads Recommendation and Auction Scoring Modeling at Google}

\author{
  \IEEEauthorblockN{George Kurian$^*$, Somayeh Sardashti$^*$, Ryan Sims, Felix Berger, Gary Holt, Yang Li, Jeremiah Willcock}
  \IEEEauthorblockN{Kaiyuan Wang, Herve Quiroz, Abdulrahman Salem, Julian Grady}
  \IEEEauthorblockA{
    Google LLC \\
    \{gkurian, somayeh, rwsims, flx, gholt, yangliyl, jewillco, kaiyuanw, hqz, asalem, jpg\}@google.com
  }
}

\maketitle
\pagestyle{plain}

\begin{abstract}
Large-scale Ads recommendation and auction scoring models at Google scale demand immense computational resources. While specialized hardware like TPUs have improved linear algebra computations, bottlenecks persist in large-scale systems. This paper proposes solutions for three critical challenges that must be addressed for efficient end-to-end execution in a widely used production infrastructure: (1) \textbf{Input Generation and Ingestion Pipeline}: Efficiently transforming raw features (e.g., "search query") into numerical inputs and streaming them to TPUs; (2) \textbf{Large Embedding Tables}: Optimizing conversion of sparse features into dense floating-point vectors for neural network consumption; (3) \textbf{Interruptions and Error Handling}: Minimizing resource wastage in large-scale shared datacenters. To tackle these challenges, we propose a \textit{shared input generation} technique to reduce computational load of input generation by amortizing costs across many models. Furthermore, we propose \textit{partitioning}, \textit{pipelining}, and \textit{RPC (Remote Procedure Call)} coalescing software techniques to optimize embedding operations. To maintain efficiency at scale, we describe novel \textit{preemption notice} and \textit{training hold} mechanisms that minimize resource wastage, and ensure prompt error resolution. These techniques have demonstrated significant improvement in Google production, achieving a 116\% performance boost and an 18\% reduction in training costs across representative models.
\end{abstract}

\def\thefootnote{*}\footnotetext{Authors contributed equally to this work}\def\thefootnote{\arabic{footnote}}

\section{Introduction}
\label{sec:introduction}

Our industry-scale production infrastructure trains thousands of Ads recommendation \cite{acun2021understanding}\cite{anil2022factory} and auction scoring \cite{mcmahan2013ad} models annually across numerous teams at Google on a large-scale fleet of Tensor Processing Units (TPUs) \cite{norrie2021tpu}\cite{jouppi2023tpu}. Several  of these production models contain billions of parameters, and train on petabytes of input data, across billions of training events. Increasingly numerous machine learning (ML) components are deployed in such models to serve high quality ads at scale.

Machine learning (ML) components are computationally intensive, demanding significant compute (FLOPS), memory, and network bandwidth resources. Over time, the models have grown in size and complexity, leading to high capital expenses and power costs. ML-centric hardware accelerators like TPUs and GPUs have emerged to reduce the cost to train such large models.  These accelerators  contain large matrix units to efficiently execute linear algebra computations \cite{norrie2021design}\cite{chetlur2014cudnn}. For example, there are several  generations of TPUs (available on Google Cloud) to train ML models with increasingly higher performance and perf/Watt.

While new accelerators offer significant potential to boost ML compute, maintaining peak performance and utilization at production scale presents substantial challenges. This paper optimizes large-scale ML training on new accelerators by focusing on three key areas:

\textbf{(1) Input Generation and Ingestion:} To optimize the utilization and cost-efficiency of expensive TPU resources in a production pipeline, we need to focus on maximizing their throughput. A crucial aspect of this is the input preprocessing stage, which prepares streams of data for TPU computation. It is essential to design a fast input pipeline that minimizes downtime and ensures a continuous flow of data to the TPUs. Input feature transformation is a critical stage in the training input pipeline that converts raw features (e.g., “search query”) into numeric inputs to inject to the model’s input layer. This process, often dominated by complex dataflow and irregular matching (e.g., string processing), does not execute efficiently on conventional accelerators. Hence, they require auxiliary compute resources such as CPUs, RAM, and disk. In production scale, the cost of such auxiliary resources is a significant portion of overall training cost. To minimize end-to-end training cost and maximize performance, this paper proposes a scalable input pipeline, including a shared input generation service that reduces input generation load by amortizing the cost across several training pipelines, and a horizontally scalable per-model input reader service to keep TPU efficiency high.

\textbf{(2) Large Embedding Tables:} Embeddings are crucial in Ads recommendation and auction scoring models due to their memoization capabilities, cost-effectiveness in serving, and ability to meet strict latency requirements \cite{anil2022factory}\cite{mcmahan2013ad}.  They transform sparse feature inputs into dense floating-point vectors, which are then used by the neural network. Industry-scale models have large embedding tables which exhibit irregular and data dependent memory access patterns. Each model has hundreds of embedding tables. Across five representative production model families, the embedding tables are diverse in nature with widely varying distributions of sizes (vocabulary size: 64 to 300M, embedding dimension: 1 to 380), and arithmetic intensities (feature values per example: 1 to 230). This highlights the complexity and variability inherent in embedding models, underscoring the need for specialized optimization techniques. TPUs have hardware support for embedding primitives \cite{jouppi2023tpu}. However, due to the large  scale of embedding tables in production models, limited memory capacity in accelerators, and variability inherent in embedding tables, effectively partitioning embedding table parameters, and associated computation across multiple chips is a complex NP-hard challenge. To address these challenges, this paper proposes novel partitioning strategies for embeddings. Furthermore, it proposes pipelining, and RPC (Remote Procedure Call) coalescing software techniques to optimize performance. While mapping neural network models such as Multi-Layer Perceptrons (MLPs) and transformers onto accelerators have been extensively studied in the literature, mapping and optimizing embedding models has received much less attention.

\textbf{(3) Minimizing Wasted Resources:} In large-scale machine learning (ML) services, minimizing the idle time of expensive hardware accelerators like TPUs is crucial for cost-effective training. Ensuring uninterrupted operation, prompt error resolution, and resource wastage minimization are key. In industry-scale ML services, models are often trained in large-scale shared datacenters, managed by cluster managers [6], that house the necessary computational resources (e.g., TPUs, CPUs, storage). However, these environments can experience job preemptions due to priority changes, software updates, or hardware maintenance. To maintain high system efficiency in such scenarios, this paper explains techniques to prevent resource wastage due to preemptions and to efficiently resolve errors caused by both permanent conditions (e.g., numeric overflow due to model divergence) or transient conditions (e.g., network interruptions). By addressing these challenges, the proposed methods optimize resource utilization of large-scale ML training in shared datacenter environments.

Finally, Ads models require continuous online training to adapt to fast-changing user preferences and advertiser inventories. In addition to training on petabytes of historical data, these models must train on recent data continuously to maintain model freshness \cite{anil2022factory}\cite{zhang2024wukong}. From a systems perspective, training on recent data has much lower training speed requirements than training on historical data. While historical training must consume several years worth of training data in a few days, recent data can be consumed only at the rate it is produced. This difference has implications on the system size (i.e., number of chips) required to train the model, and the optimizations deployed. Our production scale system accounts for such variations, optimizing performance when training on both historical and recent data.

This paper presents Google's centralized platform for Ads models that addresses the challenges of large-scale Ads recommendation and auction scoring models. This paper makes the following contributions to improve the efficiency of large-scale Ads model training on TPUs:
\begin{enumerate}
\item We present the details of Google's centralized platform for training Ads models with heterogeneous model architectures and highly variable resource requirements. This dedicated infrastructure executes and monitors a large number of continuous training pipelines, while deploying performance optimizations with minimal intervention from users.
\item We detail the significant enhancements made to our training infrastructure that have led to the current state-of-the-art performance for production training.
\item We propose a scalable input pipeline, including a shared input generation service that reduces input generation load by amortizing the cost across several training pipelines, and a horizontally scalable per-model input reader service to keep TPU efficiency high.
\item We introduce partitioning, pipelining, and RPC coalescing techniques to optimize embedding table lookup operations on both TPUs and CPUs.
\item We describe a novel preemption notice mechanism that minimizes resource wastage caused by interruptions from other jobs in the same shared datacenter.
\item We propose a training hold infrastructure that intelligently pauses models encountering permanent errors, preventing wastage of TPU resources.
\end{enumerate}

Section \ref{sec:background} provides background on the model architectures, and TPU hardware used for training. Section \ref{sec:training_service_architecture} details the training system architecture and the distributed jobs involved. Section \ref{sec:input_generation_and_ingestion} focuses on the input generation framework and scaling techniques for heterogeneous model architectures. Section \ref{sec:orchestrating_embeddings} explores novel optimization techniques for sparse embedding operations. Section \ref{sec:training_efficiency} addresses system efficiency improvements that prevent resource wastage. Section \ref{sec:evaluation} evaluates the proposed techniques and system optimizations. Section \ref{sec:lessons} details the lessons learned and future directions. Section \ref{sec:related_work} discusses related work, and Section \ref{sec:conclusion} concludes the paper.

\section{Background}
\label{sec:background}

This section first describes the model architectures of recommendation models, and then provides a brief background on the TPUs used for training.

\subsection{Target Model Architecture}
At a 10,000-foot view, Ads recommendation and auction scoring models are composed of an embedding layer followed by several dense (hidden) neural network layers. Inputs to such models are generally categorical features (e.g., words in a search query). Embeddings are the standard way to transform categorical features into dense feature vectors for training the dense network \cite{mikolov2013efficient}. Embeddings are implemented using lookup tables. For example, consider a table with 10,000 rows (vocabulary size) and 100 columns (output dimensions). Each input feature can look up either one row (univalent embedding) or a small, dynamic number of rows (multivalent embedding, typically combined by a simple aggregation function such as an average). A neural network model might have several tables of various sizes for different categorical features. Embeddings are a key component of Ads recommendation models, and typically form the first layer, and the bulk of parameters.

The dense layers capture higher level feature information and interactions used to predict the task at hand, e.g. click-through and conversion rates in recommendation models. The standard form of dense layers is fully connected layers. The output neurons are a weighted average of the input neurons, followed by a nonlinear activation function. More recently, transformers \cite{vaswani2017attention} and mixture-of-experts models \cite{shazeer2017outrageously} have been used to improve the prediction quality. Dense layers are characterized by a large number of multiply-accumulate (MAC) operations.

\subsection{TPUs}

Tensor Processing Units (TPUs) are application specific integrated circuits (ASICs) designed by Google to accelerate machine learning workloads. TPUs were used for training recommendation models starting with TPU v2 \cite{norrie2021design}, and improved in performance and perf/Watt with TPU v3 and TPU v4 \cite{jouppi2023tpu}. Each TPU chip is connected to high bandwidth memory (HBM) (16 GiB on TPUv2, 32 GiB on TPUv3/ TPUv4), and to each other through a dedicated high-bandwidth 2-D/3-D torus network called the inter-chip interconnect (ICI).  TPUs can be configured for different system sizes. These sizes range from 1 to 3072 chips on TPU v4.

TPUs consist of two types of cores: (1) TensorCore, and (2) SparseCore. The TensorCore contains wide systolic arrays and vector units that accelerate matrix multiply, and vector-based operations on large tensors. Such operations are typical in fully connected, transformer, and mixture-of-experts layers. The SparseCore is designed to accelerate embedding lookups and updates. Embedding operations are characterized by small gather/scatter memory accesses, and variable length all-to-all data exchange. Each SparseCore has multiple independent smaller cores with vector processing units (VPUs), and access to ICI with high bisection bandwidth to accelerate embedding operations. Since embedding tables are large and consume hundreds of gigabytes to terabytes, they are partitioned across the memory of many TPU chips.

\section{Training Service Architecture}
\label{sec:training_service_architecture}

Our training service comprises two key components: (1) a fleet of distinct training pipelines, each dedicated to an individual  model, and (2) a shared feature transformation infrastructure utilized by multiple models, as shown in Figure \ref{fig:adbrain_components}. Each training pipeline includes: (a) Trainer jobs, (b) a Controller, (c) Input Readers, and (d) Checkpoint Files stored in a distributed disk-based file system. Google's Borg cluster manager \cite{verma2015large} oversees all jobs within the infrastructure.
\subsection{Initial vs Caught-Up Training}

Model training is organized into two distinct phases: (a) initial training on historical training data and (b) caught-up training on data as it streams in. During initial training, models train on petabytes of historical input data, across billions of training events collected over many months. Minimizing training time is of paramount importance to developer velocity. We train across many TPU chips, typically 64-256, to complete this training phase within a couple of days. 

Once a model completes initial training, it is considered \textit{caught-up} since it has trained on all events until the present, and the model transitions to caught-up training. During this phase, the model only trains on newly arriving data. We can use a small number of TPU chips, typically 1–4, during this training phase, to stay caught-up.

\subsection{Trainer Jobs}

Trainer jobs, the core of the training pipeline, consist of TPU Host Workers and optional Parameter Servers \cite{dean2012large}. TPU host workers reside on servers directly connected to the TPU chips, primarily responsible for supplying training data to TPU chips, but otherwise performing little actual computation. A special primary TPU host worker is additionally responsible for orchestrating the TensorFlow (TF) training loop.

During initial training, the aggregate memory of all TPU chips suffices to store large embedding tables. We place embeddings in TPU memory with the tables partitioned across chips. TPU chips then handle both computationally intensive and memory intensive tasks. TensorCores on TPU Chips take care of computations and update model weights, while embeddings (including applying embedding gradients) are handled by SparseCores.

During caught-up training,  the limited TPU slice size (e.g., 1–4 chips) may necessitate offloading embedding tables to Parameter Servers. These parameter servers are regular Jobs equipped with sufficient CPU and RAM resources to accommodate embeddings. TPU host workers then handle the interactions with the Parameter Servers, sending/receiving RPCs to read and update embeddings. This can significantly impair training speed. We describe optimizations to improve speed in Section \ref{sec:orchestrating_embeddings}.

\begin{figure}
\centering
\includegraphics[width=\linewidth]{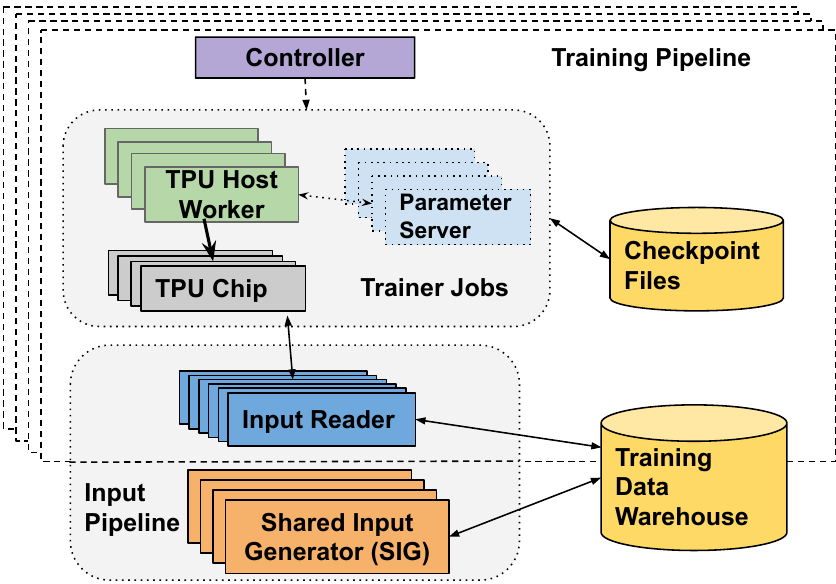}
\caption{High level components of the training service, with corresponding dataflow.}
\label{fig:adbrain_components}
\end{figure}

\subsection{Controller}

The Controller is in charge of orchestrating the training cycles called epochs. An epoch is some duration of wall time (e.g., 1 hour) during which a model trains on a range of input events. During training, TPU host workers request work units (groups of training events) from the Controller. The Controller decides what data to train on, including events not yet seen by the model or previous events with new labels.

At the end of the epoch, the Controller writes a checkpoint. To ensure the checkpoint is complete and consistent, the Controller forces all pipeline components to drain outstanding work and quiesce. This forced stall ensures state is consistent across the training pipeline. If any training component fails for any reason, the Controller does not write a new checkpoint. The next epoch then begins by restoring from the most recent checkpoint and retraining on the data from the failed epoch. Since the checkpointing protocol ensures complete and consistent pipeline state at the end of every successful epoch, the model will train on all data exactly once, and no training data is lost or duplicated.

\section{Input Generation and Ingestion}
\label{sec:input_generation_and_ingestion}

A key to efficient, production-scale model training is optimizing data preparation and ingestion. While TPUs excel at handling the computationally expensive mathematical operations of training, the feature transformations needed to convert raw data into numerical model inputs are also resource-intensive, but not amenable to hardware acceleration. If feature computation and data ingestion lag behind graph execution on TPUs, valuable accelerator resources are wasted.

To overcome this bottleneck, we designed our input pipeline with two primary components: a Shared Input Generator (SIG) for reading and pre-processing raw data, and input readers that supply input batches to trainer jobs (Figure \ref{fig:adbrain_components}). This architecture ensures a seamless flow of data, maximizing TPU utilization and accelerating model training.

\subsection{Shared Input Generation (SIG): Optimizing Feature Transformation for Model Training}

\begin{figure}
\centering
\includegraphics[width=\linewidth]{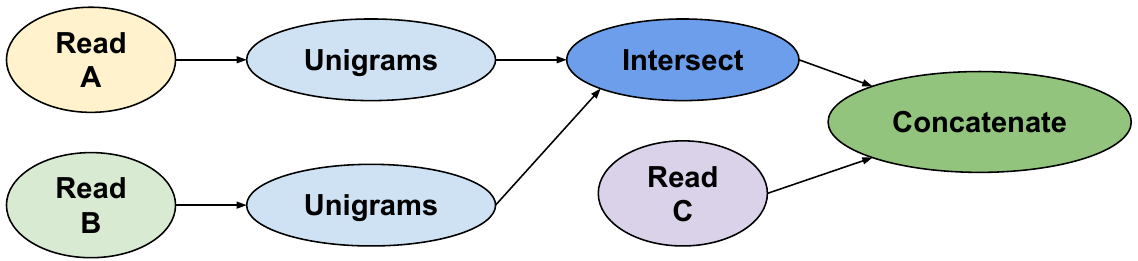}
\caption{Example connected component from an Ads recommendation model’s input transformation graph.}
\label{fig:sig_example}
\end{figure}

In our platform, models use an in-house domain-specific language (not Tensorflow) to describe input features through transformation graphs. Each input to the model consumes the output of one connected component in the graph; each component shows how to read, combine and transform raw data to produce the input. For instance, consider a model input that requires reading text fields A and B, converting them to unigrams, computing their intersection, and concatenating the result with another text field C (Figure \ref{fig:sig_example}).

The sheer scale of our system, with individual models consuming petabytes of raw data and the collective processing of trillions of events per second, necessitates a focus on high throughput. While models rarely share their entire transformation graph, individual connected components are frequently reused. To leverage this commonality and optimize resource utilization, we introduce Shared Input Generation (SIG). This mechanism materializes the output of shared components, effectively amortizing the computational cost across multiple models.

\begin{figure}
\centering
\includegraphics[width=\linewidth]{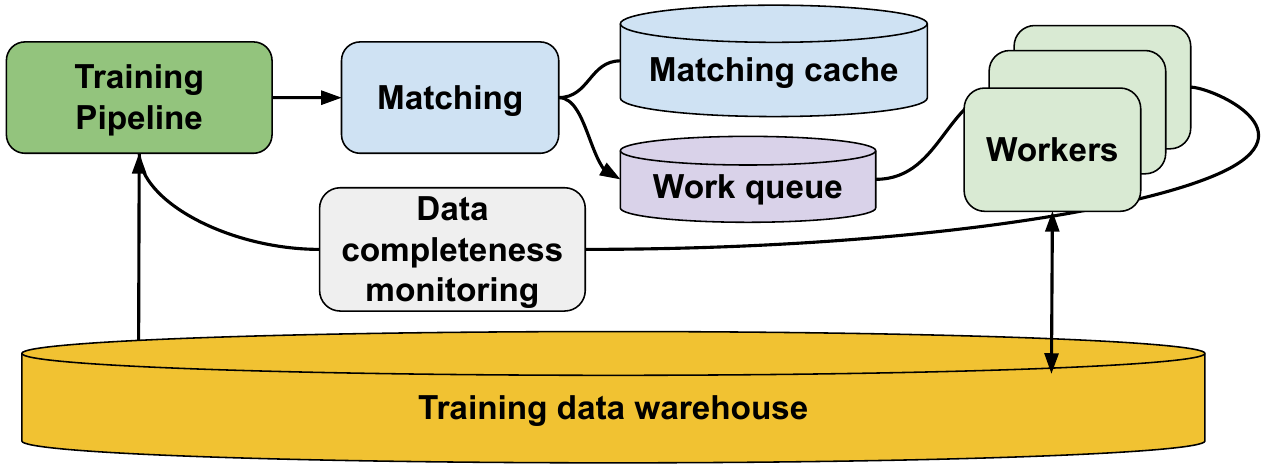}
\caption{System design and architecture of the shared input generation (SIG) pipeline.}
\label{fig:sig_workflow}
\end{figure}

Figure \ref{fig:sig_workflow} outlines SIG's high-level design:
\begin{enumerate}
\item Training models submit their transformation graphs to SIG's Matching frontend.
\item The frontend returns a "read solution" to the model's training pipeline, specifying which pre-computed (memoized) data to fetch from the general training data warehouse.
\item This warehouse integration minimizes the need for custom logic to distinguish between memoized and raw data access.
\end{enumerate}

By strategically caching and reusing intermediate results, SIG ensures efficient feature transformation, even when dealing with massive datasets and high-throughput demands. This approach significantly reduces redundant computations, accelerating model training and optimizing overall system performance.

To determine which data a training pipeline should read (the "read solution"), SIG performs a two-step process:
\begin{enumerate}
\item \textbf{Matching}: The transformation graph submitted by the pipeline is compared against components stored in SIG's Matching Cache. This cache holds the components of all transformations that SIG has previously memoized (pre-computed and stored). The comparison assesses both the structure and content of the graphs. A cache hit signifies that SIG has performed the necessary transformations, allowing reuse of the memoized data. Conversely, a cache miss indicates that some or all of the pipeline's transformations haven't been materialized yet.
\item \textbf{Scheduling New Work}: For cache misses, SIG schedules new memoization tasks. It assigns a storage location for the output of these tasks, includes it in the read solution, and enqueues the tasks in a global queue. Consequently, the read solution comprises a combination of existing memoized data and data that will be generated by upcoming tasks. Training pipelines can track the progress of these tasks and access the results as they become available using out-of-band completeness monitoring infrastructure provided by the training data warehouse.
\end{enumerate}

\textbf{SIG Workers}, the backbone of the memoization process, continuously monitor the task queue, fetch raw data, apply transformations, and store results back in the warehouse. By utilizing a distributed disk-based storage system, the warehouse accommodates datasets far exceeding RAM capacity. These workers are globally replicated and strategically positioned alongside the training data to maximize read throughput. This approach ensures efficient handling of massive datasets, as SIG can pre-compute and store common transformation results, minimizing redundancy and speeding up model training.

\subsubsection{Challenges and Solutions} Despite its effectiveness, SIG faces several key challenges:

\textbf{Scaling}: The dynamic nature of SIG's workload poses a significant challenge in resource management. SIG workers must rapidly scale up to meet the demands of TPU training jobs, often experiencing a tenfold increase in capacity within hours. Conversely, they need to scale down promptly when memoized results are readily available to avoid wasting resources. This rapid fluctuation in resource requirements—with CPU usage potentially surging by a factor of 100 within hours—creates a complex environment for resource planning and monitoring. To tackle this challenge, ongoing efforts are focused on optimizing resource allocation and forecasting in this volatile landscape.

\textbf{Prioritization}: Within the model training landscape, not all pipelines carry the same weight. Some, like those developing high-revenue-impact models, naturally take precedence over more exploratory or research-oriented ones. Additionally, these priorities are not static; they evolve as models are created, refined, and retired.
To manage this dynamic prioritization, SIG employs a client-based approach. Each model belongs to a specific "client," and each client has its own dedicated work queue, which constantly adapts to the changing landscape of models and their respective priorities. The priority assigned to each work element within the queue directly reflects the priority of the model that requested it. This client-centric approach ensures that resources are allocated efficiently, with higher-priority models receiving preferential treatment. It allows for a flexible and responsive system that can adapt to changing priorities during model deployment.

\textbf{Insertion}: SIG's current strategy involves materializing (pre-computing and storing) all requested data. This non-adaptive insertion policy prioritizes system simplicity over potential efficiency gains that could be achieved with a more sophisticated, adaptive approach. The effectiveness of the current policy is evident in the high cache hit rates, consistently exceeding 95\%. Moreover, the average data block stored in the cache is utilized by 22 different models, demonstrating the substantial reuse of memoized results across various training pipelines. This data suggests that even without an adaptive insertion policy, SIG effectively leverages shared computations to accelerate model training.

\textbf{Eviction}: Managing storage costs is crucial for the sustainability of SIG. To achieve this, a dedicated eviction process constantly monitors the "last requested" time, which is stored alongside each entry in the matching cache. This allows SIG to identify and remove memoized data that hasn't been used recently, freeing up valuable storage space. In addition, there are cases where data needs to be evicted for other reasons. For example, users might discover a bug in a transformation graph and request the removal of all memoized outputs associated with that graph. This situation becomes particularly complex when the problematic graph is a subgraph shared across multiple transformation graphs. To address this, SIG incorporates a search infrastructure that indexes transformation graphs based on various attributes, including the raw features they read, the training pipelines that utilize them. We provide a web-based UI where users can query the system, precisely pinpoint and evict all memoized data generated by a faulty subgraph, ensuring data integrity and preventing the propagation of errors in downstream processes.

\subsubsection{Limitations and Future Directions} Despite its crucial role in enabling large-scale TPU training, SIG has notable limitations that present opportunities for future enhancements:

\begin{enumerate}
\item \textbf{Limited Memoization Scope}: SIG currently only memoizes connected components within transformation graphs, not arbitrary subgraphs. This restriction hinders optimizations like eliminating common nested subgraphs across different models.
\item \textbf{Inability to Join Memoized and Raw Data}: The input pipeline lacks the capability to combine memoized SIG data with raw feature data. Consequently, SIG cannot selectively memoize only the most impactful transformations.
\item \textbf{Unsupported Mutable or Late-Arriving Data}: SIG does not handle mutable or late-arriving data, which is common in caught-up training scenarios. This limitation arises from the fact that late-arriving data can be altered by delayed processing (e.g., spam detection), and incorporating such mutable data would complicate SIG's amortization logic. As a workaround, training pipelines needing mutable data switch to the less efficient Local Input Generation (LIG).
\item \textbf{Large Failure Blast Radius}: The dependency of multiple training pipelines on a single SIG worker pipeline amplifies the impact of failures. In the current implementation, SIG workers often process multiple components simultaneously, leading to cascading failures if a single component is misconfigured.
\end{enumerate}

We acknowledge these limitations and are actively working towards addressing them in a future version of SIG.

\subsection{Input Reading: Adapting to the Demands of High-Performance TPU Architectures}

The increasing power of newer TPU architectures has created a challenge: TPU hosts lack sufficient CPU resources to directly read and transform raw data from the data warehouse. To address this, we developed input readers, a horizontally scalable system of stateless jobs tailored to each training pipeline. This design enables flexible adaptation to varying workloads.

Horizontally scaling the input readers allows the system to adapt to the changing needs of the processing workload. For instance, during the transition from initial training to caught-up training, the workload shifts from being I/O-bound to compute-bound. Input readers seamlessly handle this change, performing feature transformations entirely within themselves during the caught-up phase. This process, known as Local Input Generation (LIG), differs from SIG as it does not share transformations across models.

\textbf{Stateless Readers}: Input readers are designed to be stateless, ensuring flexibility and scalability. TPU host workers (Figure \ref{fig:adbrain_components}) request work units from the Controller which delegates them to input reader tasks. These reader tasks stream training examples (events) to the TPU host, where they are buffered in memory. The host then constructs input batches from this buffer.

\textbf{Horizontal Auto-scaling}: Horizontal scaling is achieved by adjusting the number of in-flight work units requested by TPU hosts. This is done dynamically by monitoring the fullness of the host's buffer. If events are consumed faster than they are added, the host requests more work units, increasing parallelism and, in turn, scaling up input reader tasks.

\section{Orchestrating Embeddings}
\label{sec:orchestrating_embeddings}

As discussed in Section \ref{sec:background}, there are two kinds of model components, embeddings and dense layers. The computation and communication patterns of dense layers, and embeddings are quite different. Dense layers (fully connected and transformers) focus on matrix and wide vector operations that are regular, statically shaped and independent of the input data. Operations on embeddings focus on smaller vector operations, are quite irregular, and data dependent. For our models, embeddings are large, and need to be partitioned across several machines or chips to fit. In this section, we present our approaches for TPU embeddings, when embeddings are also placed/handled by TPU chips, and for CPU embeddings which are placed on Parameter Servers. The partitioning techniques presented for TPU embeddings are applicable and generalizable to all models using TPUs (not just Ads recommendation models) since they have been incorporated within the TPU compiler.

\subsection{TPU Embeddings}
The embedding tables are partitioned across all TPUs using a hybrid of model parallelism techniques. The goal of the partitioner is to maximize performance by balancing the load on compute and memory resources while respecting all hardware-imposed constraints. Partitioning is an N-P hard problem, and solved by modeling it as a constraint optimization problem. We use both heuristics and ILP (Integer Linear Programming) solvers to solve the optimization problem. The number of nodes for partitioning is the number of SparseCores (SC) in the TPU supercomputer.

\begin{figure*}
\centering
\includegraphics[width=\linewidth]{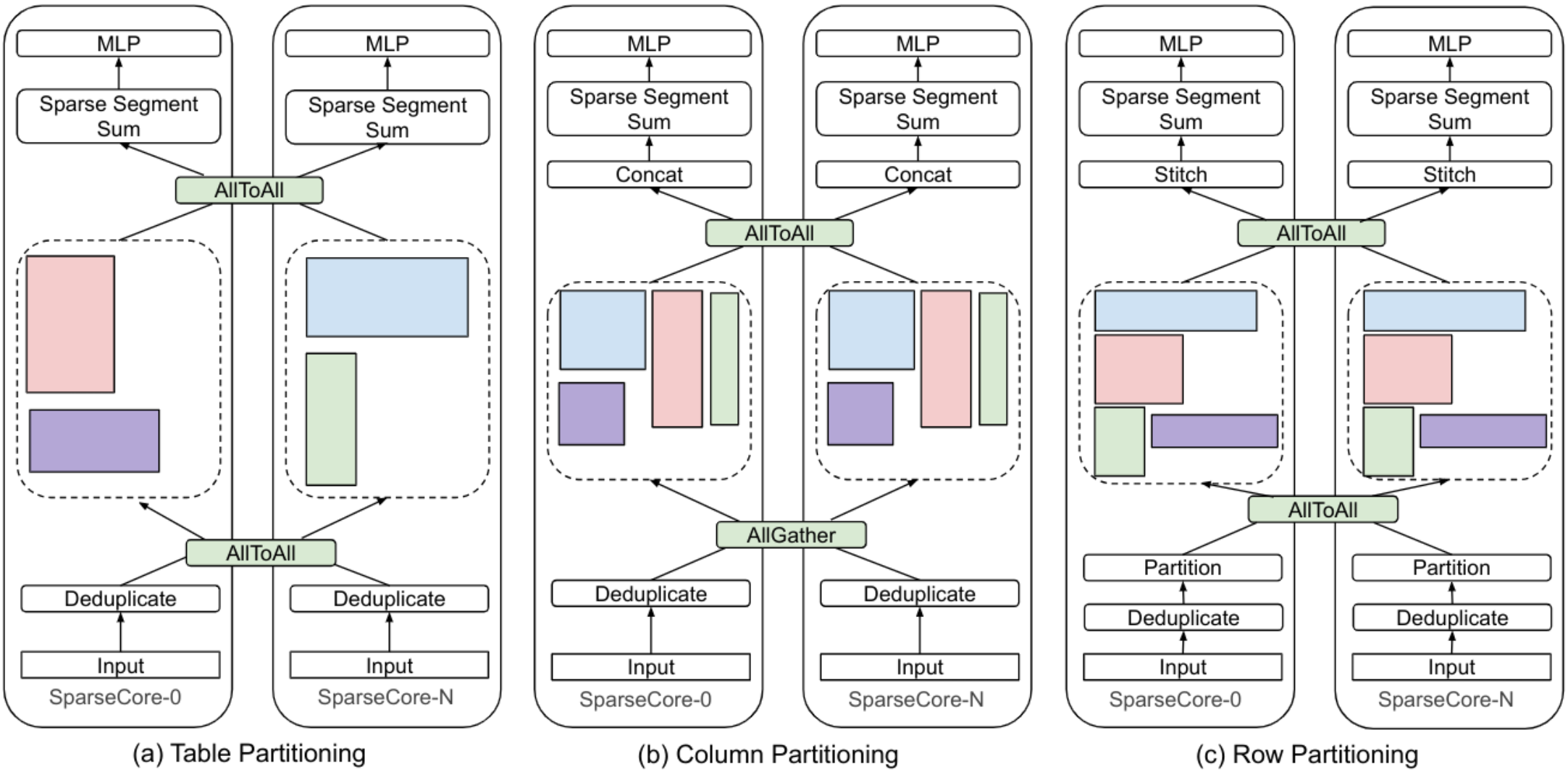}
\caption{Embedding Table Partitioning strategies for TPU. There are three model parallelism strategies used: (a) table partitioning, (b) column partitioning, and (c) row partitioning. Each color coded block is an embedding table. The communication primitives used are AllToAll and AllGather. We employ a similar schematic \cite{mudigere2022software}, to illustrate dataflow.}
\label{fig:embedding_partitioning}
\end{figure*}

There are three methods to use model parallelism for partitioning an embedding layer: table, column, and row partitioning.  All three can be used simultaneously. Figure \ref{fig:embedding_partitioning} shows how these partitioning techniques are applied in the TPU.

\textbf{Table partitioning} places different embedding tables on different nodes; it works well when the number of tables exceeds the number of nodes. The main limitation is that the embedding table vocabulary sizes and widths differ greatly across tables, stranding memory. Also, embedding tables are typically multivalent: the number of feature values per example per table (valency) is more than one. The valency is dynamic (dependent on the training event), and unknown at compile time, making load balancing a challenge.

\textbf{Column partitioning} splits tables along their width resulting in multiple smaller, sharded tables; it works well with wide tables as shards are perfectly load-balanced. However, column sharding leads to smaller and more memory accesses during embedding table lookups and updates, lowering HBM bandwidth utilization. This technique also increases feature value processing overheads since each feature value indexes multiple shards. Moreover, only element-wise training optimizers are compatible with column partitioning—optimizers that operate on an entire embedding row, e.g., LAMB \cite{you2019large}, SM3 \cite{anil2019memory}, don’t work.

\textbf{Row partitioning} splits tables along their vocabulary size; it works well with large vocabularies that are the bulk of memory bytes and accesses. The main limit is the need for application-level load balancing strategies. Typically, we use random hashing for input feature values to ensure equal traffic to all shards. However, if the vocabulary is sorted by access frequency, cyclic distribution \cite{distribution} (spreads the hottest rows over multiple nodes) is better than block distribution \cite{distribution} (places the hottest rows on the first node).

All model parallelism methods use all-to-all data exchange with deduplication. The input batch is deduplicated to obtain a list of unique feature values which are partitioned and exchanged with remote SCs (Figure \ref{fig:embedding_partitioning}). The embedding vectors corresponding to these unique feature values are fetched from remote HBM and a sparse segment sum \cite{tensorflow_segment_sum} operation computes the per-example embedding vectors.

In contrast, Meta’s \cite{mudigere2022software} model parallelism strategy on GPU partitions and exchanges all feature values with remote GPUs. The remote GPUs perform segment sum \cite{tensorflow_segment_sum} operations to compute the embedding vectors. With row partitioning, the segment sums computed are partial; a collective reduce-scatter operation is used to obtain the final embedding vectors. Hence, the traffic injected into the network with this scheme is proportional to the node count. Our strategy works better for our internal workloads for three reasons. First, deduplication of Zipf-distributed embedding accesses significantly reduces network traffic. Second, for row partitioning, our strategy scales network load based only on the unique feature value count in a batch but does not scale with node count. Finally, deduplication prevents hotspots for small tables, avoiding the need for data parallelism for their optimal handling. Small tables can be handled using one of the model parallelism methods without load imbalance concerns.

\subsection{CPU Embeddings}

During caught-up training, we shard embeddings across parameter servers, and communicate the embedding values and gradients over the datacenter network. TPU chips still perform the bulk of model computation, and run the forward/backward passes of training. The model training speed depends on two factors: (1) how well the parameter servers are load balanced, and (2) the amount of RPCs and network traffic incurred to the parameter servers.  

There are several methods of partitioning the embeddings. The most useful technique for CPUs is Row Sharding, where all embedding tables are split into groups of rows across all parameter servers. The advantage with this approach is that the parameter servers are well load-balanced. The drawback is that the number of RPCs incurred with this approach is large. RPCs are sent on behalf of each TPU core to the parameter servers for all embedding tables. The traffic pattern is all-to-all, and the \#-of-RPCs to each parameter server = \#-of-TPU cores × \#-of-Tables, resulting in significant scalability bottlenecks. In a production Ads model performing online training, there are hundreds of embedding tables, and tens of TPU cores. This results in thousands of RPCs from TPU host workers to each parameter server on every training step.

\textit{Fetch and update coalescing} is a technique where values from different tensors are combined into the same RPC. All communication between parameter servers and the TPU workers happens through a single pair of RPCs per batch; the values for different embeddings are coalesced into a single serialized packet. This turns out to be critical for performance. \textit{Table Stacking}, where tables with the same width and optimization parameters are stacked to form a single variable, is also critical to reduce the graph complexity. With table stacking, we require the use of cyclic distribution \cite{distribution} to ensure that each table is spread across all parameter servers for optimal load balance. 

\subsection{Software Optimization Techniques}

Beyond the partitioning techniques described above, a few software optimizations bear mentioning. 

\textbf{Feedback-directed partitioning (FDP)}. For efficient partitioning, the TPU software probabilistically profiles training batches, counting the number of feature values (per example), number of unique feature values, and load imbalance information, storing these statistics in a database shared across many models. The database is indexed by feature metadata (e.g., feature name). Most experimental models reuse features, enabling accurate prediction of embedding patterns. We sample only 3\% of the training batches to reduce training overheads. The database of statistics is continuously updated.

\textbf{Software-based deduplication}. TPUv2/v3 uses a software-based technique to identify duplicated feature values and save work. The host CPU deduplicates and passes the unique feature values to the SparseCore (SC). The SC gathers embedding vectors for each unique value, and the TensorCore (TC) performs segment sum \cite{tensorflow_segment_sum} operations to construct per-example embedding vectors. The wide range of valency (peak 1000, but averages closer to 10) further complicates deduplication. To avoid overprovisioning TC memory, the TC handles the common, low-valency case while large valencies are handled without deduplication on the SC. The profiling database from feedback directed partitioning is used to inform the shapes for operations on the TensorCore. Since TPUv4 has added the capability for full deduplication in hardware, software-based deduplication is unnecessary.

\begin{figure}
\centering
\includegraphics[width=\linewidth]{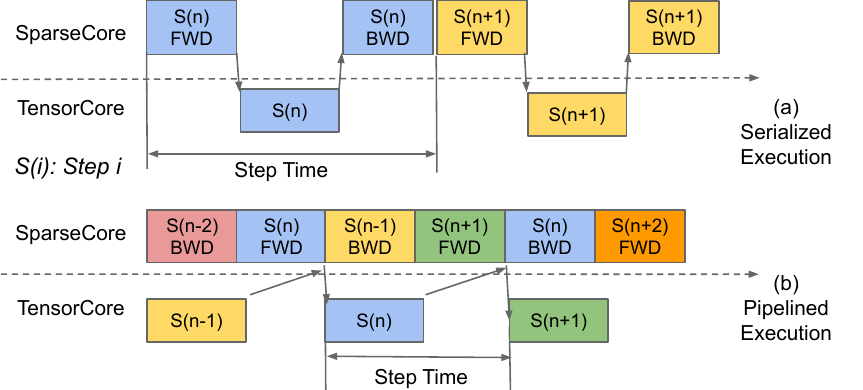}
\caption{(a) Serialized and (b) Pipelined execution of TensorCore with SparseCore. Pipelining execution helps improve overlap, and thereby performance.}
\label{fig:pipelining}
\end{figure}

\textbf{Pipelining with TensorCore}: Since embeddings are the input for recommendation models, forward embeddings are the first stage and backward embeddings are the last stage of a training step. Strictly serialized execution (Figure \ref{fig:pipelining}a) would force SC and TCs to alternate execution during training. We relax the strict serial semantics and allow the SC to start step N+1 while the TensorCore runs step N (Figure \ref{fig:pipelining}b). This relaxation better overlaps SC and TC operation, yielding significant performance improvement compared to serialized execution. Note however, that the overlap increases contention for shared resources, primarily HBM (high bandwidth memory) and ICI (inter-chip interconnect) bandwidth. The resulting embedding gradients are stale (by one step), but this was found to not impact model quality in internal recommendation and ranking models.

\section{Training Efficiency}
\label{sec:training_efficiency}

Improving system efficiency requires that work done by TPUs are not wasted due to interruptions from other jobs, and error conditions are handled promptly with minimal wastage of TPU resources.

\subsection{Handling Errors}

Training pipelines may routinely get into states where no training progress can be made. It is vital for resource efficiency to automatically detect these states and release the TPUs so they can be assigned to another queued service. There are 2 categories of states that need to be detected:

\textbf{Permanent errors}:  Invalid model configurations, compilation errors, out-of-memory errors, and numeric overflow errors (e.g., NaNs) are examples of permanent errors. When the system detects such an error, it places a training hold signaling that user intervention is required. The hold is not removed automatically.

\textbf{Transient stalls}: A primary example is unavailability of training data in SIG because SIG is still busy processing the materialization request for the event range the model trains over next. The training pipeline can detect this as part of scheduling work units and place a training hold releasing the expensive compute resources (TPUs and input readers). The Controller of the system stays up during the hold and removes it once the data has become available.

There could also be temporary errors during training caused by events like network interruptions. In such cases, training holds are not placed, and model training is restarted from the last checkpoint.

\subsection{Maintenance and Preemptions}

Training pipeline lifetimes vary from days to months. These long time lines mean that jobs and machines in the service are subject to interruptions due to higher priority jobs in the same shared datacenter, new software releases, or machine maintenance; they cause jobs to be terminated prematurely, and moved to different machines.

To better support these interruptions, training jobs support a preemption protocol. External components such as the deployment or release infrastructure can send a preemption notice and give the job a warning that it will be shut down soon. When a job receives such a notice, it informs the Controller of the impending shutdown, which in turn broadcasts the notice to all other jobs.

Each job immediately attempts to finish its outstanding work and transitions into a state where it can correctly checkpoint the training progress. Once the training progress is checkpointed, the preempted job will exit gracefully. The Controller will block progress in the next training epoch until the job has restarted and joined the training pipeline. This ensures that training progress made in the current epoch is not discarded due to interruptions.

\section{Evaluation}
\label{sec:evaluation}

In this section, we evaluate the performance and efficiency improvements from our main contributions.

\subsection{Methodology}

The system studied for initial training (on historical data) is comprised of 128 TPUv4~\cite{jouppi2023tpu} chips. Each TPUv4 machine contains 4 TPU chips, each connected to 32 GiB of High Bandwidth Memory (HBM), and 2 AMD Rome~\cite{amd_rome} CPU sockets, each connected to 256 GiB of DDR4 RAM. In addition to TPUv4 machines, each training pipeline relies on input readers and (optional) parameter servers placed on external CPU machines. Such CPU machines are shared by several heterogeneous jobs running in the datacenter.

We pick five representative recommendation models accounting for $>$50\% of our production workloads for the study. Each model has 50M to 300M dense model (MLP and transformer) weights and hundreds of embedding tables. The embedding tables are diverse in nature with widely varying distributions of sizes (vocabulary size: 64 to 300M, embedding dimension: 1 to 380), and arithmetic intensities (feature values per example: 1 to 230).

\subsection{Shared Input Generation (SIG)}

\begin{figure}
\centering
\includegraphics[width=\linewidth]{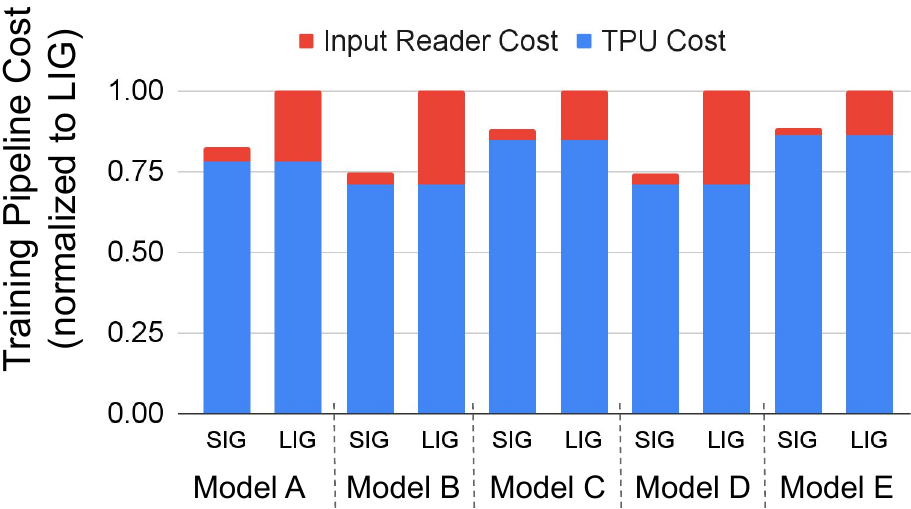}
\caption{Cost of the training service. The cost of input readers is much higher with LIG while the TPU cost is the same (assuming TPUs are not blocked by inputs).}
\label{fig:sig_results}
\end{figure}

SIG significantly improves resource utilization for the input reader job by amortizing the cost of input generation across several models. With local input generation (LIG), the input generation cost must be paid separately for every model even though there is significant overlap in the features processed by multiple models. Figure \ref{fig:sig_results} shows the normalized cost of the training pipeline for five representative recommendation models running on 128 TPU chips. The cost is defined as the total TCO (CapEx + OpEx) of the TPU machines and input readers required to train the model end-to-end (the CapEx is normalized over the lifetime of the machine). The cost includes TPU chip, CPU chip, RAM, tray, power provisioning and power delivery costs.

The input reader cost is much higher with LIG due to its higher CPU and RAM footprint. SIG reduces the input reader cost by 4.3x to 7.5x compared to LIG. The TPU cost is identical with SIG and LIG since SIG primarily makes input readers more efficient. The TPU cost dominates the training pipeline cost, hence SIG only reduces total cost by 12\% to 27\% across the five representative models: the geometric mean of the cost reduction is 18\%.

In practice, the benefits with SIG are understated in Figure \ref{fig:sig_results}. It wasn’t practical to obtain sufficient CPU and RAM resources for all LIG experiments. When external CPU/RAM resources are low, we found that TPU utilization suffered as a result of the high input reader cost in LIG. When TPU utilization is low, the model training cost with LIG is greatly increased. Hence, the benefits with SIG were in reality much higher. SIG was a necessity to keep our large TPU footprint busy.

SIG has its own pool of workers (that do not belong to a training pipeline) to actually do the input generation (as shown in Figure \ref{fig:sig_workflow}). The cost of SIG’s worker pool depends on the number of models sharing the same feature transform graph. If more models rely on the same transform graph, the SIG worker pool cost is reduced. For our models, the overall SIG CPU/RAM cost was only 4\% the combined cost of all the training services due to amortization across thousands of models actively using SIG. The SIG hit rate (i.e., transform graph reuse) was $>$95\% in our experiments over all our models. Each memoized component was used by 22 models on average, but the peak could be higher than 400.

\subsection{Input Reading}

The need for distributed input reading can be shown by comparing the CPU resources used by Input Readers with the CPU resources available on TPU host workers. Table \ref{tab:input_reading} shows the distribution of the ratio of CPU resources used for input reading to those available on TPU hosts across our training pipelines. Values $\le 1$ indicate that the input reading usage fits within the TPU host, and values $> 1$ indicate that we require more CPU resources for input reading than are available on the TPU host. From Table \ref{tab:input_reading}, the input reading resources required by 50\% of our training pipelines can be satisfied by the TPU host (ratio is $0.7$). However, if we consider the 90th percentile, input reading requires the resources of 3.5 TPU hosts. Hence, even in the optimistic case where all TPU host resources could be used solely for input reading, the host resources are not sufficient for a significant fraction of our training pipelines, motivating the need for distributed input reading.

\begin{table}[!h]
\begin{center}
\begin{tabular}{ l | c | c | c | c }
 Percentile & 50\% & 90\% & 95\% & 99\% \\ 
 \hline
 CPUs (normalized) & 0.7 & 3.5 & 6.6 & 16.0
\end{tabular}
\end{center}
\caption{Input Reader CPU usage relative to TPU host limit.}
\label{tab:input_reading}
\end{table}

The average input reader CPU usage also varies widely by training stage. During initial training, we use SIG, hence the average CPU usage is lower than caught-up training where we use LIG. During initial training, the ratio of CPUs used for the input reader to that available on the TPU host (averaged across all Ads training jobs) is 0.4 while the same ratio is 1.3 during caught-up training. This motivates the need for dynamic horizontal scaling for the input readers.

\subsection{Embedding Optimizations for TPU}
\begin{figure}
\centering
\includegraphics[width=\linewidth]{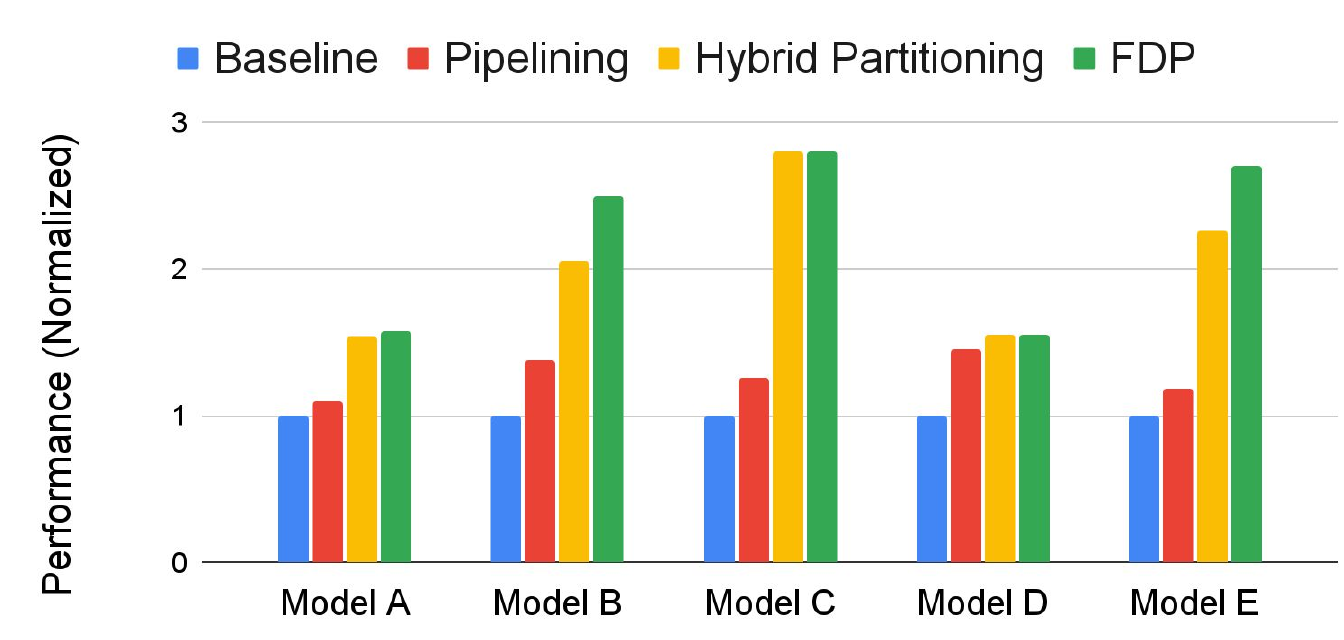}
\caption{Impact of embedding optimizations on the performance of recommendation models.}
\label{fig:embedding_results}
\end{figure}

This subsection discusses the various improvements for TPU embeddings (Figure \ref{fig:embedding_results}). Recall that the dense computations in the recommendation model (MLPs, transformers) are handled by the TensorCore, and the embeddings are handled by the SparseCore. \textbf{Baseline} is the performance without any optimizations. TensorCore/SparseCore pipelining is disabled, we employ a simplistic partitioning scheme using only row partitioning, and there is no feedback directed partitioning. \textbf{Pipelining} is the performance after enabling TensorCore/SparseCore pipelining. \textbf{Hybrid Partitioning} is the performance after further enabling row, column, and table partitioning. \textbf{FDP} is the performance after further enabling feedback directed partitioning. All models benefit from most optimizations.

\textbf{Baseline}: In this scheme, the activations are computed for the dense layers on the TensorCore only after embedding lookups on the SparseCore are complete. This is done so as to respect graph dependencies and ensure mathematical correctness (recall that the embedding layer is the first layer in recommendation models). Similarly, the activation gradients are computed for the dense layers before embedding updates on the backward pass are performed. The TensorCore and SparseCore are not overlapped for the majority of the training step. The partitioning scheme only involves row sharding.

\textbf{Pipelining}: When pipelining is enabled, the TensorCore and SparseCore execution are completely overlapped. The resulting training step time is the maximum of the TensorCore and SparseCore step times. However, note that the TensorCore and SparseCore operations both become slower due to contention for HBM (high bandwidth memory), ICI (inter-chip interconnect), and shared on-chip resources. Moreover, if either the TensorCore or SparseCore consume the majority of step time in the Baseline scheme, pipelining will not significantly improve performance.

From Figure \ref{fig:embedding_results}, Pipelining improves the performance of all models. Models A and E benefit to a smaller extent than other models since they are significantly SparseCore bound without further partitioning improvements (the SparseCore consumes the majority of the step time in Models A and E since the time taken to process embeddings is larger than that required to execute the dense computation).

From a model quality perspective, pipelining causes the embedding gradients to become stale (by one step). We conducted extensive offline and live experiments to determine whether this impacts the final trained model accuracy. The conclusion was there was no noticeable quality impact in all models studied.

\textbf{Hybrid Partitioning} further improves the performance of all models through better load balancing of the work done across SparseCores. Due to inherent load imbalance across embedding rows (even with cyclic distribution), it is difficult to achieve uniform load distribution using only row partitioning. Enabling table and column partitioning of embedding tables further improves load balance across SparseCores. The final partitioning scheme is a hybrid mix of table, column and row partitioning.

To illustrate why table and column partitioning improves performance, let's use a simplified example. Consider a model with two embedding tables, $T_1$ and $T_2$. Each table is composed of 4 rows and 64 columns. The model is partitioned on a TPU system with 4 SparseCores ($SC_0$, $SC_1$, $SC_2$, $SC_3$). For both tables, the average number of embedding lookups to each row (after deduplication) is 0.6 ($Row_0$), 0.3 ($Row_1$), 0.2 ($Row_2$), and 0.1 ($Row_3$) respectively. The average number of lookups is computed using a statistical mean across many training steps. The number of lookups to each row is different due to the variance in frequencies of the corresponding feature values in the input training data.

We define a term called \textit{Load Imbalance} to understand the partitioning efficiency. To calculate load imbalance, we compute the total number of bytes $B_i$ accessed from memory on each SparseCore $SC_i$ (averaged across many steps). Load imbalance is defined as the ratio of max $B_i$ to average $B_i$. Lower the load imbalance, higher is the efficiency of partitioning ($N$ is the number of SparseCores in the following equation).

\begin{equation}
    Load\ Imbalance = \frac{N\max_i B_i}{\sum_i B_i}
\end{equation}

With row partitioning, each table row is placed on one SparseCore. The load imbalance is $2$ ($= \frac{4 \times 0.6}{0.6 + 0.3 + 0.2 + 0.1}$). If table and column partitioning are leveraged instead of row partitioning, each table would be partitioned along its width into two segments, for a total of four segments. Each segment would be placed on one SparseCore. More concretely, $T_0$ would be placed on ($SC_0$, $SC_1$), and $T_1$ would be placed on ($SC_2$, $SC_3$). Columns $0$-$31$ would be placed on ($SC_0$, $SC_2$), and columns $32$-$63$ would be placed on ($SC_1$, $SC_3$). The resulting load imbalance factor is $1$, leading to better efficiency than using row partitioning.

In addition to load imbalance, HBM bandwidth utilization also plays a significant role in performance. Column partitioning reduces HBM bandwidth utilization by lowering the memory access granularity. Instead of fetching the contents of an entire embedding row from memory, column partitioning fetches several partial rows instead. This leads to lower bandwidth utilization but improved load imbalance. In practice, the hyper-parameters for hybrid partitioning are selected by modeling it as a constraint optimization problem. The optimization goal is minimizing the execution time for embedding lookups and updates, and the constraints are hardware-imposed (e.g., memory capacity). Both memory access granularity and load imbalance are factors that impact execution time.

\textbf{FDP (Feedback Directed Partitioning)} improves performance by taking into account profiling information from batches (total number of feature values and unique feature values per training event) during partitioning. FDP is primarily important due to multivalent lookups. The valency for such lookups is data dependent and not known at compile time. Valency can be calculated only using statistical information gathered at runtime. FDP facilitates sharing of such information across models, and using it for efficient partitioning. FDP improves the performance of models B and E by 21\% and 19\% respectively. The speedup for model A is comparatively lower since two tables contribute to most of the embedding lookups in model A, and we are able to achieve uniform distribution of these tables even without FDP. Models C and D are TensorCore bound after hybrid partitioning, hence do not further benefit from FDP.

Overall, the embedding optimizations improve performance by 58\% to 180\% across the five representative models: the geometric mean of the improvement is 116\%.

\subsection{Embedding Optimizations for CPU}

Fetch and update coalescing improves the performance of caught-up models by 8\% (Models A, D) and 10\% (Model B, C) and 6\% (Model E). To improve the efficiency of such coalescing, we need to minimize the number of memory copies in software. This is done by implementing smart split operations for the coalesced buffer using reference counting techniques. The results of the split operation share the same underlying buffer as the original coalesced tensor, thereby avoiding memory copies.

\subsection{Handling Errors}

Table \ref{tab:chip_demand} shows the TPU chip demand for models averaged over a 7-day duration relative to the chip footprint. If $x$ chips are used for active training at any given time, another $1.03x$ chips are being demanded by models in the training queue that are ready to run but don’t have the available resources, and $2.49x$ chips are being demanded by models placed on training hold. Given the large TPU demand from models placed on hold due to encountering permanent errors or transient stalls, it is vital to handle error conditions carefully to avoid stranding (under-utilizing) TPU resources.

\begin{table}[!h]
\begin{center}
\begin{tabular}{ l | c | c | c }
 TPU Chip Demand & Training & Queued & On Hold \\ 
 \hline
  & 1 & 1.03 & 2.49
\end{tabular}
\end{center}
\caption{TPU chip demand relative to the available ceiling.}
\label{tab:chip_demand}
\end{table}

\subsection{Maintenance \& Preemptions}
Training pipelines are subject to interruptions due to (1) higher priority jobs in the same shared datacenter, (2) new software releases, or (3) machine maintenance. During such interruptions, an early epoch end is initiated to save training progress. 61\% of the preempted epochs succeed in committing their training progress.
\section{Lessons Learned and Future Directions}
\label{sec:lessons}

While TPUs and similar hardware excel at linear algebra computations, maximizing their utilization in a production environment requires a holistic approach. Bottlenecks can arise in other stages. In this section, we highlight key lessons learned in optimizing TPU performance for large-scale machine learning systems, focusing on continuous training of Ads models:

\subsection{Preventing TPU Starvation: The Importance of Input Pipeline Efficiency}

The input pipeline, responsible for preparing and feeding data to the TPU, can significantly impact overall training performance and cost. A slow or inefficient pipeline can starve the TPU, leading to underutilization and wasted resources. For our production system, we built a scalable input pipeline with two key components: Shared Input Generation Service to reduce redundant computations by amortizing the cost of input generation across multiple models; and Horizontally Scalable Input Reader Service to ensure a continuous flow of data to the TPUs, maximizing their efficiency.

\subsection{Managing Large Embedding Tables with TPU SparseCores is Key}

Industry-scale models deal with massive embedding tables, each with millions or even hundreds of millions of entries. These tables also have diverse characteristics in terms of size and access patterns, making optimization complex. Accessing embedding entries involves irregular memory lookups, inefficient on hardware accelerators designed for structured computations. While TPUs provide special purpose hardware for handling embedding tables (SparseCore in TPUs), partitioning large embedding tables within the limited memory capacity of accelerators like TPUs is a significant challenge. To address this, we came up with Novel Partitioning Strategies to effectively distribute embedding tables across multiple accelerator chips, and Pipelining and RPC Coalescing techniques to further enhance performance by streamlining embedding operations and minimizing communication overhead.

\subsection{Robust Resource Management Strategy is Crucial}

Hardware accelerators like TPUs are expensive to operate. Minimizing their idle time is essential for cost-effective training, especially at scale. In shared datacenters, training jobs can be interrupted due to various reasons like priority changes, software updates, or hardware maintenance. This can lead to significant resource wastage if not handled properly.
Training processes can also encounter errors, both permanent (e.g., model divergence) and transient (e.g., network issues). Efficiently resolving these errors is crucial to prevent prolonged downtime. In this paper, we addressed these issues by Preemption Notices to anticipate and handle preemptions gracefully, and Efficient Error Resolution to quickly identify and address both permanent and transient errors, ensuring prompt recovery and minimal downtime.

\subsection{Continuous Training Is Challenging To Scale on TPUs}

Effectively adjusting TPU slice sizes and embedding placements to match the varying computational needs of different training phases is challenging. While training on historical data requires high speed to process years of data quickly, recent data training only needs to match the rate of data generation. The difference in speed requirements affects the necessary system size (number of accelerator chips) for each type of training. In this paper, we discussed how we handle such training phases (caught-up training vs initial training) to balance training speed, resource utilization, and overall cost.

\subsection{Future Directions}

There are two main directions that are being explored to improve the efficiency of Shared Input Generation. (1) \textbf{Storage overhead reduction}: Memoizing arbitrary subgraphs that are expensive to compute vs the full input transformation graph is helpful in reducing the storage overhead. In addition, joining memoized data with raw feature data that is cheap to compute can provide additional improvements. (2) \textbf{Handling mutable data}: This would allow caught-up training to use SIG, boosting auxiliary resources savings. However, note that in our Ads recommendation system, revenue-critical models would all be trained on this mutable data; memoizing and sharing data between such models introduces shared-fate production risks that are complex to solve.

To avoid tight coupling of embedding storage space and compute throughput, hybrid partitioning techniques are being explored. This involves placing the most frequently accessed tables or table rows in TPU HBM, and placing less frequently accessed ones on external variable servers. The bookkeeping overhead to do this classification is the most challenging aspect, especially in the face of changing feature distributions for embedding tables.

\section{Related Work}
\label{sec:related_work}

In this paper, we explore the intricacies of using TPUs to train large-scale Ads recommendation and auction scoring models at Google \cite{anil2022factory}. Internet companies widely employ such recommendation models \cite{acun2021understanding}\cite{naumov2019deep} in their datacenters. As data scales rapidly, training such large models faces many challenges \cite{wang2022rec}. The rapid innovation in ML models has made ML accelerators like GPUs and TPUs increasingly prevalent in the industry. For example, Meta's MTIA \cite{firoozshahian2023mtia} is specifically designed for inference workloads, particularly in deep learning recommendation models, while Google's TPU v4 \cite{jouppi2023tpu} is a recent domain-specific architecture optimized for both training and serving these models. 

The industry and research communities have adopted various system-wide techniques, such as distributed training frameworks and data pipelining, to address the challenges of deploying accelerators for complex, large-scale models. For example, \cite{acun2021understanding} has highlighted the limited support for handling large embedding tables and intensive feature transformation with massive input data as key challenges when using GPUs for training. In response, software-hardware co-design solutions have been proposed to enhance scalability \cite{mudigere2022software}, and improve embedding table partitioning across GPUs. FlexShard \cite{sethi2023flexshard} improves the performance of sequence based embedding tables by sorting rows based on their access probability, replicating the frequently accessed rows, and row partitioning the less frequently accessed ones. In our partitioning method, deduplication effectively provides this benefit by reducing the number of accesses to frequently accessed rows. RecShard \cite{sethi2022recshard} uses mixed integer linear programming (MILP) to partition embedding tables across multiple GPUs and memory tiers (HBM vs UVM) taking into account statistical feature distributions. Ads recommendation models could benefit from using this technique during caught-up training on TPUs, by sharding embedding tables across both TPU HBM and external variable servers based on the statistical distribution of accesses to a table.

Prior research has also delved into continuous integration of model checkpoints \cite{liu2022monolith}, online learning techniques for GPU-based CTR models \cite{xu2021agile}, and strategies for managing large embeddings in distributed systems \cite{lian2022persia}. Additional insights have been published on online training systems at industry scale and the importance of maintaining model freshness \cite{jiang2019xdl}\cite{he2014practical}.

The data storage and ingestion pipeline is another critical aspect, with the input pipeline often identified as a major bottleneck in training performance \cite{zhao2022understanding}\cite{zhang2024wukong}. Approaches such as central data warehouses, data pre-processing services, and extract-transform-load (ETL) jobs have been proposed \cite{zhao2022understanding}. Furthermore, ML data processing services with automatic scaling \cite{graur2022cachew}, frameworks for caching feature transformations \cite{phani2022uplift}, and data preprocessing services built on TensorFlow’s tf.data \cite{audibert2022case} have been explored to enhance efficiency and reduce costs. Nectar \cite{gunda2010nectar} presents several techniques, including caching, for managing data processing at scale, while \cite{derakhshan2022materialization} introduces materialization and reuse mechanisms to eliminate redundant data processing in machine learning pipelines.

\section{Conclusion}
\label{sec:conclusion}

This paper presents optimizations to improve the end-to-end performance, and reduce training costs for Ads recommendation model training at Google. The contributions include the following: (1) a scalable input pipeline, including a shared input generation service that reduces input generation load by amortizing the cost across several training pipelines, and a horizontally scalable per-model input reader service to keep TPU efficiency high, (2) partitioning, pipelining, and RPC coalescing techniques to optimize embedding table lookup operations on both TPUs and CPUs, (3) a novel preemption notice mechanism that minimizes resource wastage caused by interruptions from other jobs in the same shared datacenter, and (4) a training hold infrastructure that intelligently pauses models encountering permanent errors, preventing wastage of TPU resources. The techniques described improve performance by 116\%, while simultaneously lowering training cost by 18\% by reducing the reliance on external CPUs and RAM.

\section*{Acknowledgements}

This work would not have been possible without the foundational efforts of several people who worked on multiple generations of this product, including Igor Tsvetkov, Keith Arner, and Evan Benshetler. Their insights and dedication laid the groundwork for the advancements presented here.

\bibliographystyle{IEEEtranS}
\bibliography{references}

\end{document}